\documentstyle[aps,twocolumn]{revtex}
\begin{document}
\draft
\title{
Level statistics in a metallic sample: Corrections to the
Wigner--Dyson distribution.
}
\author{Vladimir E.Kravtsov$^1\dagger$ and Alexander D. Mirlin$^2\ddagger$}
\address{ $^1$ International Centre for Theoretical Physics,
P.O. Box 586, 34100 Trieste, Italy}
\address{
$^2$ Institut f\"{u}r Theorie der Kondensierten Materie,
  Universit\"{a}t Karlsruhe, 76128, Karlsruhe, Germany}
\date{June 23, 1994}
\maketitle
\begin{abstract}
Deviation of the level correlation function in a mesoscopic metallic
sample from the Wigner--Dyson distribution is calculated by using a
combination of the renormalization group and non-perturbative treatment.
For given spatial dimension the found correction is determined by
 the sample conductance.

\end{abstract}
\pacs{PACS numbers: 71.30.+h, 05.60.+w, 72.15.Rn}
\narrowtext

The problem of level correlation in quantum systems attracts interest of
physicists since the work of Wigner \cite{wig}.
The random matrix theory developed
by Wigner and Dyson \cite{dys} describes well the level statistics
of various complex systems, like nuclei or atoms.
Later on, Gor'kov and Eliashberg \cite{GE} put forward an assumption
that  the random matrix theory is also applicable to the problem
of energy level correlations of
a quantum particle in a random potential. This hypothesis was proven
by Efetov, who showed \cite{efe}
that the level--level correlation function $R(\omega)$ (its formal
definition is given below) is described by the Wigner--Dyson distribution
for $\omega\ll E_c$, $E_c$ being the Thouless energy. For $\omega\gg E_c$,
the behavior of the correlation function changes  because the
corresponding time scale $\omega^{-1}$ is smaller than the time $E_c^{-1}$
the particle needs to diffuse through the sample. The form of the
correlation function in this region is dependent on spatial dimensionality
and was calculated in Ref.\cite{shkl} by means of the diffusion perturbation
theory.

In the present Letter we find a correction to the Wigner--Dyson distribution
in the region $\omega\sim\Delta\ll E_c$, $\Delta$ being the mean level
spacing. This is not a trivial task, because we calculate a correction to the
result which is essentially non--perturbative.

We study the two--level correlation function $R(s)$ defined as
\begin{equation}
R(s)={1\over \langle\nu\rangle^2}\langle\nu(E)\nu(E+\omega)\rangle
\label{eq1}
\end{equation}
where $s=\omega/\Delta$, $\nu(E)$ is the density of states at the energy $E$
and $\langle\ldots\rangle$ denote averaging over realizations of the random
potential. As was shown by Efetov \cite{efe}, the correlator (1) can be
expressed
in terms of a Green function of certain supermatrix $\sigma$--model.
Depending on whether the time reversal and spin rotation symmetries
are broken or not, one of three different $\sigma$--models is relevant, with
unitary, orthogonal or symplectic symmetry group. We will consider the case
of the unitary symmetry (corresponding to the broken time reversal
invariance) throughout the paper; the results for two other cases will be
presented at the end. The expression for $R(s)$ in terms of the
$\sigma$--model then reads:
\begin{eqnarray}
&&R(s)=\left({1\over 4V}\right)^2 \mbox{Re} \int DQ(\bbox{r})
\left[\int d^d\bbox{r}\mbox{Str} Q\Lambda k\right]^2     \nonumber\\
&&\exp\left\{-{\pi\nu\over 4}\int d^d\bbox{r} \mbox{Str} [-D(\nabla Q)^2
-2i\omega\Lambda Q]\right\}      \label{eq2}
\end{eqnarray}
Here $Q=T^{-1}\Lambda T$ is $4\times 4$ supermatrix, with $T$
 belonging to the coset space $U(1,1|2)$,
$\Lambda=diag\{1,1,-1,-1\}$, $k=diag\{1,-1,1,-1\}$,
Str denotes the supertrace,  $V$ is the
system volume and $D$ is the classical diffusion constant. To find the
detailed description of the model and  mathematical entities involved,
a reader is refered to Refs.\cite{efe,zirn}.

For $\omega\ll E_c\sim D/L^2$ ($L$ being the system size, so that
$V=L^d$) the leading contribution to the integral (\ref{eq2}) is
given by the spatially uniform fields $Q(\bbox{r})=Q$. Then
the functional integral in eq.(\ref{eq2})
reduces to an integral over a single supermatrix $Q$ and can be calculated
yielding the Wigner--Dyson distribution \cite{efe}:
\begin{equation}
R_{WD}(s)=1-\frac{\sin^2(\pi s)}{(\pi s)^2}
\label{eq3}
\end{equation}
The aim of the present paper is to calculate a correction to eq.(\ref{eq3})
due to spatial fluctuations of $Q(\bbox{r})$ in eq.(\ref{eq2}). The procedure
we are using for this purpose is as follows. We first decompose $Q$ into the
constant part $Q_0$ and the contribution $\tilde{Q}$ of higher modes with
non--zero momenta. Then we use the renormalization group ideas and
integrate out all fast modes. This can be done {\it perturbatively}
provided the dimensionless conductance $g=E_c/\Delta\gg 1$. As a result,
we get an integral over the matrix $Q_0$ only, which has to be calculated
{\it non-perturbatively}. We believe that this combination of the
perturbative renormalization--group--type and non-perturbative treatment
is of a methodological interest and might be used for other applications.

To begin with, we present the correlator $R(s)$ in the form
\begin{eqnarray}
&& R(s)={1\over (2\pi i)^2} {\partial^2\over \partial u^2}\int DQ
\exp\{-{\cal F}(Q)\}|_{u=0}\ ;  \nonumber\\
&& {\cal F}(Q)=-{1\over t}\int \mbox{Str}(\nabla Q)^2
+\tilde{s}\int \mbox{Str}\Lambda Q +\tilde{u}\int\mbox{Str} Q\Lambda k
\label{eq4}
\end{eqnarray}
where $1/t=\pi\nu D/4$, $\tilde{s}=\pi s/2iV$, $\tilde{u}=\pi u/2iV$. Now we
decompose $Q$ in the following way
\begin{equation}
Q(\bbox{r})=T_0^{-1}\tilde{Q}(\bbox{r})T_0
\label{eq5}
\end{equation}
where $T_0$ is a spatially uniform matrix and $\tilde{Q}$ describes all modes
with non-zero momenta.
When $\omega\ll E_c$, the matrix $\tilde{Q}$ fluctuates only weakly near the
origin $\Lambda$ of the coset space. In the leading order,
$\tilde{Q}=\Lambda$, thus reducing (\ref{eq4}) to a zero-dimensional
$\sigma$--model, which leads to the Wigner--Dyson distribution (\ref{eq3}).
To find the corrections, we should expand the matrix $\tilde{Q}$ around
the origin $\Lambda$. This expansion starts as \cite{efe}
\begin{equation}
\tilde{Q}=\Lambda\left(1+W+{W^2 \over 2}+\ldots\right)\ ,
\label{eq5a}
\end{equation}
where $W$ is a supermatrix with the following block structure:
\begin{equation}
W=\left(\begin{array}{ll} 0 & t_{12} \\ t_{21} & 0 \end{array}\right)
\label{eq5b}
\end{equation}
Substituting this expansion into eq.(\ref{eq4}), we get
\begin{eqnarray}
&&{\cal F}={\cal F}_0+{\cal F}_1+O(W^3)\ ;\nonumber\\
&&{\cal F}_0=\int\mbox{Str}\left[{1\over t}(\nabla W)^2+\tilde{s}Q_0\Lambda+
\tilde{u}Q_0\Lambda k\right]               \nonumber\\
&&{\cal F}_1={1\over 2}\int\mbox{Str}[\tilde{s}U_0\Lambda W^2 +
\tilde{u}U_{0k}\Lambda W^2]
\label{eq6}
\end{eqnarray}
where $Q_0=T_0^{-1}\Lambda T_0$,
$U_0=T_0\Lambda T_0^{-1}$, $U_{0k}=T_0\Lambda k T_0^{-1}$.
Let us define ${\cal F}_{eff}(Q_0)$ as a result of elimination of the
fast modes:
\begin{equation}
e^{-{\cal F}_{eff}(Q_0)}=e^{-{\cal F}_0(Q_0)}\langle e^{-{\cal F}_1(Q_0,W)}
\rangle_W
\label{eq7}
\end{equation}
where $\langle\ldots\rangle_W$ denote the inetgration over $W$. Expanding
up to the order $W^4$, we get
\begin{equation}
{\cal F}_{eff}={\cal F}_0+\langle {\cal F}_1\rangle -
{1\over 2}\langle{\cal F}_1^2\rangle + {1\over 2} \langle{\cal F}_1\rangle^2
+\ldots
\label{eq8}
\end{equation}
The integral over the fast modes can be calculated now using the Wick theorem
and the contraction rules \cite{efe,AKL}:
\begin{eqnarray}
&&\langle\mbox{Str} W(\bbox{r})PW(\bbox{r'})R\rangle=\Pi(\bbox{r}-\bbox{r'})
\nonumber\\ &&\qquad\times
(\mbox{Str}P\mbox{Str}R-\mbox{Str}P\Lambda\mbox{Str}R\Lambda)\ ;\nonumber\\
&&\langle\mbox{Str}[W(\bbox{r})P]\mbox{Str}[W(\bbox{r'})R]\rangle=
\Pi(\bbox{r}-\bbox{r'})
\mbox{Str}(PR-P\Lambda R\Lambda);\nonumber\\
&&\Pi(\bbox{r})=\int{d^d q\over (2\pi)^d} \frac{\exp(i\bbox{qr})}{\pi\nu Dq^2}
  \label{eq9}
\end{eqnarray}
where $P$ and $R$ are arbitrary supermatrices.
The result is:
\begin{eqnarray}
&&\langle{\cal F}_1\rangle=0\ ;\nonumber\\
&&\langle{\cal F}_1^2\rangle=
{1\over 2}\int d\bbox{r}d\bbox{r'} \Pi^2(\bbox{r}-\bbox{r'})
(\tilde{s}\mbox{Str}Q_0\Lambda+\tilde{u}\mbox{Str}Q_0\Lambda k)^2
\label{eq10}
\end{eqnarray}
Substituting eq.(\ref{eq10}) into eq.(\ref{eq8}), we find
\begin{eqnarray}
&&{\cal F}_{eff} (Q_0)={\pi \over 2i}s\mbox{Str}Q_0\Lambda+{\pi\over 2i} u
\mbox{Str} Q_0\Lambda k \nonumber \\
&&+{a_d\over 16 g^2} (s\mbox{Str} Q_0\Lambda +
u\mbox{Str}Q_0\Lambda k)^2  \ ; \nonumber\\
&& a_d={1\over \pi^4}
\sum_{\begin{array}{c}n_1,\ldots,n_d=0\\n_1^2+\ldots +n_d^2>0
\end{array}}^{\infty}{1\over (n_1^2+\ldots+n_d^2)^2}
\label{eq11}
\end{eqnarray}
Using now eq.(\ref{eq4}), we get the following expression for the
correlator to the $1/g^2$ order:
\begin{eqnarray}
&&R(s)=\mbox{Re}{1\over (2\pi i)^2}\int dQ_0\left\{\left({\pi\over 2i}\right)^2
(\mbox{Str} Q_0\Lambda k)^2 \right.\nonumber\\
&&\times[1-{a_d\over 16 g^2}s^2 (\mbox{Str} Q_0\Lambda)^2]
\nonumber \\ &&
\left.-{a_d\over 8g^2} (\mbox{Str} Q_0\Lambda k))^2 +
{\pi a_d\over 8 g^2 i}s(\mbox{Str}Q_0\Lambda)(\mbox{Str}Q_0\Lambda k)^2
\right\} \nonumber \\
&&\times\exp\{-{\pi\over 2i}s\mbox{Str}Q_0\Lambda\}
\label{eq12}
\end{eqnarray}
This integral over the supermatrix $Q_0$ can be calculated by using the known
technique \cite{efe}, yielding
\begin{equation}
R(s)=1-\frac{\sin^2(\pi s)}{(\pi s)^2}+\frac{a_d}{\pi^2 g^2}\sin^2(\pi s)
\label{eq13}
\end{equation}
The last term in eq.(\ref{eq13}) just represents the correction
of  order $1/g^2$ to the Wigner--Dyson distribution. The formula
(\ref{eq13}) is valid for $s\ll g$. Let us note that  the
smooth (non-oscillating) part of this correction
in the region $1\ll s\ll g$ can be found by using purely perturbative approach
\cite{shkl,shkl1}. For $s\gg 1$ the leading perturbative contribution
to $R(s)$ is given by a two--diffuson diagram:
\begin{eqnarray}
R_{pert}(s)-1 &=& {\Delta^2 \over 2 \pi^2} \mbox{Re} \sum_{
\begin{array}{c}
q_i=\pi n_i/L \\ n_i=0,1,2,\ldots
\end{array} }
{1\over (Dq^2-i\omega)^2} \nonumber\\
&=& {1\over 2\pi^2}\mbox{Re}\sum_{n_i\ge 0}
{1\over (-is+\pi^2g\bbox{n}^2)^2}
\label{eq15a}
\end{eqnarray}
At $s\ll g$ this expression is dominated by the $\bbox{q}=0$ term, with
other terms giving a correction of order $1/g^2$:
\begin{equation}
R_{pert}(s) = 1-{1\over 2\pi^2 s^2} + {a_d\over 2\pi^2 g^2}\ ,
\label{eq15b}
\end{equation}
where $a_d$ was defined in eq.(\ref{eq11}). This formula is obtained in the
region $1\ll s\ll g$ and is perturbative in both $1/s$ and $1/g$. It does
not contain oscillations (which cannot be found perturbatively) and gives no
information about actual small--$s$ behavior of $R(s)$. The result
(\ref{eq13}) of the present Letter is much stronger: it represents the exact
(non--perturbative in $1/s$) form of the correction in the whole region
$s\ll g$.

The calculation presented above can be straightforwardly generalized to the
other symmetry cases. The result can be presented in a form valid for all
three cases:
\begin{equation}
R_\beta(s)=\left(1+\frac{a_d}{2\beta\pi^2 g^2} \frac{d^2}{ds^2}s^2\right)
R_\beta^{WD}(s)
\label{eq14}
\end{equation}
where $\beta=1(2,4)$ for the case of orthogonal (unitary, symplectic) symmetry;
$R_\beta^{WD}$ denotes the corresponding Wigner--Dyson distribution,
 explicit form of which can be found e.g. in \cite{efe,mehta}.
( We denote by $g$
 the conductance per one spin projection: $g=E_c/\Delta=\nu D L^{d-2}$, without
multiplication by factor 2 due to the spin.)

For $s\to 0$ the Wigner--Dyson distribution has the following behavior:
\begin{eqnarray}
&& R_\beta^{WD}\simeq c_\beta s^{\beta};\qquad  s\to 0 \nonumber\\
&& c_1={1\over 6};\
c_2={1\over 3};\ c_4={1\over 135}
\label{eq15}
\end{eqnarray}
As is clear from eq.(\ref{eq14}), the found correction does not change
the power $\beta$, but renormalizes the prefactor $c_\beta$:
\begin{equation}
R_\beta(s)=\left(1+\frac{(\beta+2)(\beta+1)}{2\beta} \frac{a_d}{\pi^2 g^2}
\right) c_\beta s^\beta\ ;\quad s\to 0
\label{eq16}
\end{equation}

As was already noted, the above derivation of eq.(\ref{eq14}) is valid
provided $g\gg 1$, i.e. in the case of a good metal. The level statistics is
also of big interest in the critical point $g=g_*$ corresponding to the
Anderson metal--insulator transition. The ``tail'' of this distribution
at $s\gg g$ was studied in the recent paper \cite{KLAA}, but nothing
has been known analytically about its behavior at $s\lesssim g$.
In $d=3$ the critical point corresponds to the strong disorder:
$g_*\sim 1$, where the Wigner--Dyson distribution loses its validity.
However, we can consider formally the spatial dimension to be
$d=2+\epsilon$ with $\epsilon\ll 1$, where $g_*\sim 1/\epsilon\gg 1$, i.e.
the transition point is still in the weak disorder regime. (one may hope that
the obtained results will be still qualitatively valid when continued to
$\epsilon$=1.) Then the Wigner--Dyson distribution has again a
parametrically broad range of validity $s<g_*$, and a question about the
corrections to it due to spatial fluctuations is meaningful. We can repeat
basically the same calculation, as was done above for the $d=3$ metallic
system. However, the diffusion propagator $\Pi(\bbox{r})$ should now
be replaced by its form in the critical region \cite{weg}:
\begin{equation}
\Pi(\bbox{r})\propto\int\frac{d^d q}{(2\pi)^d}\frac{\exp(i\bbox{qr})}
{g_* q^d}
\label{eq17}
\end{equation}
As a result, we find in close analogy with eq.(\ref{eq13}),
\begin{equation}
\delta R(s)\propto \frac{1}{g_*^2}\sin^2(\pi s)
\label{eq18}
\end{equation}
It should be noted that in contrast to eq.(\ref{eq13}), the exact numerical
coefficient in eqs.(\ref{eq17}), (\ref{eq18}) is not known. Eq.(\ref{eq18})
is written for the unitary symmetry case $\beta=2$; the general expression
for the correction is given by eq.(\ref{eq14}), with $g_*$ substituted for
$g$.

As is seen from eqs.(\ref{eq16}), (\ref{eq18}), the found corrections
lead to an enhancement of the prefactor $c_\beta$ in the low--$s$ behavior
$R(s)\approx c_\beta s^\beta$ of the correlator, that  means physically
a weakening of the level repulsion.
The correction to the prefactor $c_\beta$ is of order $1/g^2$.
When a system approaches the metal--insulator transition in $d=3$, the
conductance $g$ tends to a value $g_*$, which is of order unity. The
calculation carried out in the present paper is no more justified in this
region. We expect, however, that being formally applied it gives
qualitatively correct behavior of $R(s)$ at small values of $s$, namely
an enhancement of the coefficient $c_\beta$ by a factor of order unity.

In conclusion, we have calculated the deviation of the level--level
correlation function $R_\beta(s)$ in a mesoscopic metallic sample from its
universal Wigner--Dyson form, using the supersymmetric sigma--model approach.
The method of calculation combines a perturbative elimination of fast
spatial modes (in spirit of renormalization group ideas) and subsequent
non--perturbative evaluation of an integral over the zero mode. The found
correction is of order $1/g^2$, where $g$ is the dimensionless conductance.
It does not change the power $\beta$ of the $s^\beta$ behavior of the
correlator $R(s)$ as $s\to 0$, but renormalizes the corresponding prefactor.

V.E.K. appreciates discussions with A.G.Aronov and I.V.Lerner.
A.D.M. thanks Y.V.Fyodorov for critical reading the manuscript.
A.D.M. is grateful to the International Centre for Theoretical Physics
and the Institute for Nuclear Theory at the University of Washington
for the kind hospitality and to the Alexander von Humboldt Foundation
for financial support.

\end{document}